\documentclass[journal]{IEEEtran}
\usepackage{graphicx}
\usepackage{amsmath}
\usepackage{amssymb}
\usepackage{amsbsy}

\begin{document}
%
\title{Residual Excitation Skewness for Automatic Speech Polarity Detection}
%

\author{Thomas Drugman, Member IEEE}      

\markboth{IEEE Signal Processing Letters}%
{Shell \MakeLowercase{\textit{et al.}}: Bare Demo of IEEEtran.cls for Journals}

\maketitle

\begin{abstract}
Detecting the correct speech polarity is a necessary step prior to several speech processing techniques. An error on its determination could have a dramatic detrimental impact on their performance. As current systems have to deal with increasing amounts of data stemming from multiple devices, the automatic detection of speech polarity has become a crucial problem. For this purpose, we here propose a very simple algorithm based on the skewness of two excitation signals. The method is shown on 10 speech corpora (8545 files) to lead to an error rate of only 0.06\% in clean conditions and to clearly outperform four state-of-the-art methods. Besides it significantly reduces the computational load through its simplicity and is observed to exhibit the strongest robustness in both noisy and reverberant environments.
\end{abstract}


%

\let\thefootnote\relax\footnotetext{
\\T. Drugman is supported by FNRS, and with TCTS Lab, University of Mons, Belgium. \textit{Address:} 31 Boulevard Dolez, 7000 Mons, Belgium. \emph{Phone:} +3265374749. \emph{Email:} thomas.drugman@umons.ac.be.}

\section{Introduction}

When a microphone is used to record a speaker, inverting its electrical connections will cause an inversion of the polarity of the acquired speech signals. The origin of a polarity in the speech signal stems from the asymmetric glottal waveform exciting the vocal tract resonances. During the production of voiced sounds, the glottal source exhibitis periodic discontinuities at particular moments called Glottal Closure Instants (GCIs) \cite{Drugman-GCI}. As described by models of the glottal source (e.g \cite{SourceModels}), the speech polarity is defined as being positive if the glottal flow derivative exhibits a negative peak at the GCI. Otherwise it is said to be negative.


Our human ear is mostly insensitive to a polarity change \cite{Sakaguchi}. Nonetheless the performance of several speech processing techniques can be severely deteriorated if the signal polarity is erroneous. For example, in concatenative speech synthesis \cite{NUU}, segments of recorded speech are appended to each others. If two segments with different polarities are concatenated, this will result in a phase jump at the jointure and lead to an audible glitch. This effect will noticeably degrade the quality of the synthesized voice, in particular if it occurs in voiced segments with a relatively high energy \cite{Sakaguchi}. This will be also the case for most of the pitch-synchronous analysis and synthesis methods, such as the popular pitch-synchronous overlap-add (PSOLA) approach \cite{TDPSOLA}.

Contrarily to the techniques which rely on the magnitude component of the Fourier spectrum (as done with the traditional LPCs or MFCCs), methods dealing with the phase of the speech signal are polarity-dependent. That is the case of the great majority of the algorithms developed in glottal analysis \cite{Drugman-Thesis}. In this way, most of the methods for GCI detection, or for glottal flow estimation, parameterization and modeling, assume a positive polarity.

There is thus a large number of speech processing techniques whose performance can be dramatically affected if polarity is inverted. In parallel, these same techniques are developed to automate systems which are nowadays expected to work properly on huge amounts of data. These data can be acquired through a variety of microphones, and consequently with different polarities. Detecting correctly the speech polarity is then a necessary step to ensure the good behaviour of the aforementioned techniques.


In this paper, a new simple algorithm for automatic speech polarity detection is proposed. This method, called Residual Excitation Skewness (RESKEW), is based on the skewness of two excitation signals: the traditional residual signal, and a rough approximation of the glottal flow derivative. Contrarily to state-of-the-art approaches, it has the advantage of not requiring voicing decisions or F0 estimation. It therefore allows a fast computation, and is shown through our experiments to provide the best results in clean conditions as well as in noisy and reverberant environments.

The paper is structured as follows. Section \ref{sec:Existing} briefly describes existing techniques for polarity detection, while the proposed RESKEW method is presented in Section \ref{sec:Proposed}. Our experimental protocol is detailed in Section \ref{sec:Protocol} and the results are discussed in Section \ref{sec:Results}. Finally Section \ref{sec:Conclu} concludes the paper. 

\vspace{-4pt}


\section{Existing Methods}
\label{sec:Existing}

To the best of our knowledge, only four methods were proposed in the literature to automatically detect the speech polarity. They are here briefly described.

\vspace{-4pt}

\subsection{Gradient of the Spurious Glottal Waveforms (GSGW)}

The idea of the GSGW \cite{GSGW} technique is to investigate the discontinuity at the GCI in an estimated glottal waveform, whose sign is dependent upon the speech polarity. A criterion is proposed based on a sharp gradient of the spurious glottal waveform near the GCI \cite{GSGW}. Each glottal cycle is assigned a polarity, and the final decision considers a majority voting.

\subsection{Phase Cut (PC)}

In the PC method \cite{PC}, it is searched for the time instant where the two first harmonics are in phase (and which should correspond roughly to the GCI). As their phase slopes are linked by a factor 2, the phase value where they intersect is $\phi_{cut}=\phi_2 - 2 \phi_1$, where $\phi_1$ and $\phi_2$ denote the phase of the first and second harmonics. A value of $\phi_{cut}$ close to 0 (respectively $\pi$) is expected to be due to a positive (respectively negative) peak in the excitation \cite{PC}. Again, a majority voting is applied across voiced frames.

\subsection{Relative Phase Shift (RPS)}

The RPS technique \cite{PC} is derived from PC and exploits a greater amount of harmonics. It makes use of Relative Phase Shifts (RPS's), denoted $\theta(k)$ and defined as $\theta(k) = \phi_k - k\cdot \phi_1$, where $\phi_k$ is the instantaneous phase of the $k^{th}$ harmonic. When the excitation exhibits a positive peak, RPS's have a smooth structure across frequency. This smoothness is shown to be dramatically sensitive to a polarity inversion \cite{PC}. RPS applies a smoothness criterion on RPS's for each voiced frame, and draws the polarity decision by majority voting.

\subsection{Oscillating Moments-based Polarity Detection (OMPD)}

For a speech signal, OMPD \cite{OMPD} calculates on a sample-by-sample basis statistical moments oscillating at the local fundamental frequency. The key idea of OMPD is to compute two oscillating moments (with an odd and even orders) such that their phase shift allows to determine the correct polarity. Again, local decisions are taken for each voiced frame, and the final polarity arises from a majority voting.


\section{Proposed Technique}
\label{sec:Proposed}

\begin{figure}[!ht]
  \centering
  \includegraphics[width=0.4\textwidth]{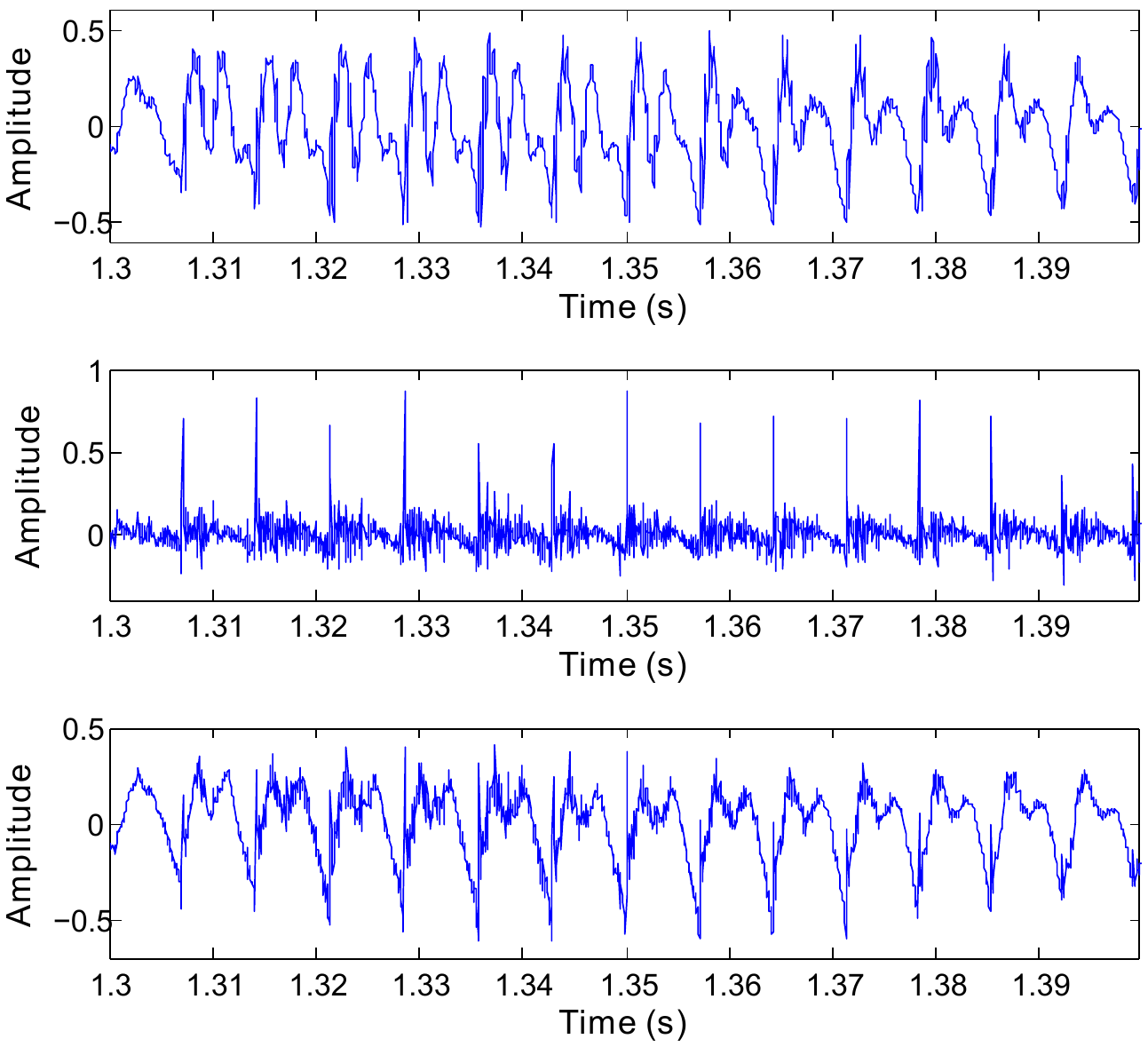}
  \vspace{-8pt}
  \caption{Example of a segment of voiced speech $s(n)$ (top panel), with its corresponding LP residual signal $r(n)$ (middle panel) and a rough approximation of its glottal flow derivative $g'(n)$ (bottom panel).}
  \label{fig:Example}
  \vspace{-12pt}
\end{figure}

The key idea of the proposed technique is that the excitation signal contains relevant information about the speech polarity, as its behaviour reflects the asymmetry of the glottal production. More precisely, two excitation signals are considered in this work. The first one is the traditional residual signal $r(n)$, obtained by estimating (here every 5 ms on 25ms-long Hanning windows) through Linear Prediction (LP) analysis the coefficients $a_i$ of an auto-regressive model of the speech signal $s(n)$, and by removing the contribution of this spectral envelope by inverse filtering. An example of a voiced speech segment together with its corresponding residual signal are displayed respectively in the top and middle plots of Figure \ref{fig:Example}. It can be noticed that the residual signal exhibits important positive peaks at GCI locations.

The second signal used by the proposed method is a rough approximation of the glottal flow derivative, denoted $g'(n)$. For this, we wanted to avoid the reliance on voicing decisions, F0 contour or GCI estimates as required by usual glottal flow estimation techniques. This is because we believe that the step of polarity detection is the very first one in an analysis workflow and should then be as simple as possible. The estimation of the glottal flow derivative is therefore carried out as follows. A high-passed version $\tilde{s}(n)$ of the speech signal $s(n)$ is obtained by keeping its spectral content beyond a cut-off frequency $F_c$, and its $\tilde{a}_i$ coefficients are extracted by LP analysis. In this paper, this high-pass filtering is carried out with an elliptic filter of order 9. The estimate of the glottal flow derivative $g'(n)$ is then achieved by inverse filtering the original speech signal $s(n)$ using the $\tilde{a}_i$ coefficients. The key idea behind this is to fix $F_c$ such that $\tilde{s}(n)$ contains most of the vocal tract information and leaves apart the low frequencies which mainly account for the glottal contribution. $F_c$ should therefore be fixed in between the glottal formant and the first formant. Again, the goal here is not to obtain an accurate and robust estimate of the glottal source, but rather a rough approximation exhibiting the speech polarity information. The bottom plot of Figure \ref{fig:Example} shows the resulting esimate of $g'(n)$ for the considered voiced speech segment. It is observed that this signal has a shape comparable to what is described by glottal flow models \cite{SourceModels}, and presents negative peaks at GCIs.


Since these two signals exhibit an asymmetry due to their discontinuities at GCIs, this should be reflected by computing the skewness of the distribution of their samples. For a series $x$ of $N$ values, the statistical skewness $\gamma_1 (x)$ is defined as:

\begin{equation}\label{eq:Skewness}
\gamma_1 (x) = \frac{\frac{1}{N} \sum_{i=1}^{N}{(x_i-\bar{x})^3}} {(\frac{1}{N} \sum_{i=1}^{N}{(x_i-\bar{x})^2})^{\frac{3}{2}}},
\end{equation}

where $\bar{x}$ is the mean of $x$. The skewness is known to be a measure of the asymmetry of probability density function. A positive skewness means that the tail on the right side of the distribution is longer than on its left side. As expected from Fig. \ref{fig:Example}, this will be the case for the residual signal, while the estimated glottal source will have a negative skewness. These two measures are obviously dependent on a polarity inversion.  

Throughout the remainder of this paper, RESKEW-res will refer to the algorithm determining the speech polarity as the sign of $\gamma_1(r(n))$, while RESKEW-glot will use the sign of $-\gamma_1 (g'(n))$. Since these two signals should have a skewness with an opposite sign, we also proposed a more robust technique referred to as RESKEW and which detects the speech polarity by inspecting the sign of $\gamma_1(r(n))-\gamma_1 (g'(n))$. It is worth noting that this calculation is not performed on a frame basis, but on the whole speech signal. A possible improvement might consist in focusing on voiced segments. This would be done at the expense of a reduction of the method simplicity, which is avoided here. Nevertheless the proposed technique is expected to be rather robust since amplitudes in voiced segments are much greater than in silence or unvoiced parts, and therefore prevail in the skewness calculation.

\section{Experimental Protocol}
\label{sec:Protocol}

Experiments are carried out on a large amount of data with 10 speech corpora. Several English voices are taken from the CMU ARCTIC database \cite{CMUARCTIC}, which was designed for speech synthesis purpose: AWB (male), BDL (male), CLB (female), JMK (male), KSP (male), RMS (male) and SLT (female). About 50 min of speech is available for each of these speakers. The Berlin database \cite{Berlin} is made of German emotional speech (7 emotions) from 10 speakers (5F - 5M) and consists of 535 sentences altogether. The two speakers RL (Scottish male) and SB (Scottish female) from the CSTR database \cite{CSTR}, with around 5 minutes per speaker, are also used for the evaluation. The total number of files over the 10 corpora is 8545, all sampled at 16kHz.

The performance of the proposed techniques is compared to the four state-of-the-art methods described in Section \ref{sec:Existing}, both in clean conditions and in environments degraded by an additive noise or by reverberation. To simulate noisy recordings, noise was artificially added to the original signals at various Signal-to-Noise Ratios (SNRs). Both a White Gaussian Noise (WGN) and a babble noise were considered. The noise signals were taken from the Noisex-92 database \cite{Noisex}, and were added so as to control the overall SNR without silence removal. For reverberant conditions, we considered the $L$-tap Room Impulse Response (RIR) of the acoustic channel between the source to the microphone. RIRs are characterised by the value $T_{60}$, defined as the time for the amplitude of the RIR to decay to -60dB of its initial value. A room measuring 3x4x5~m and $T_{60}$ ranging \{100, 200,~\ldots,~500\}~ms was simulated using the source-image method~\cite{Allen1979} and the simulated impulse responses convolved with the clean speech signals. Note that if the environment is known to be reverberant, a possible improvement of the proposed technique could consist in increasing the order of LP analysis in order to have a better estimate of the spectral envelope (which is affected by the convolutive noise). Nonetheless, the order of LP analysis has been standardly fixed to $F_s/1000+2$ ($F_s$ being the sampling rate) across all our experiments.

As performance metrics, we used the detection error rate averaged over the 10 databases. The rapidity of algorithms was also evaluated using the Relative Computation Time (RCT) defined as the ratio between the computation time using our Matlab implementation on a i7-2720 QM, 2.20 GHz processor, and the duration of the corresponding sound file.

Finally, note that for the state-of-the-art techniques which require an estimation of the voicing boundaries and of the F0 contour, we used the Summation of Residual Harmonics (SRH, \cite{SRH}) algorithm as it was shown to provide the best robustness in noisy recordings.


\section{Results}
\label{sec:Results}

\subsection{Influence of the Cut-off Frequency}

Since the order of the AR models has been fixed, the only remaining parameter used by our proposed methods is the cut-off frequency $F_c$ employed to get the rough approximation of the glottal source. As stated in Section \ref{sec:Proposed}, its value should be low enough so that $\tilde{s}(n)$ captures the first formant information, and sufficiently high to avoid that it integrates the glottal formant. Again, the goal here is to have a rough approximation of the glottal source, rather than a precise estimate.

\begin{figure}[!ht]
  \centering
  \includegraphics[width=0.4\textwidth]{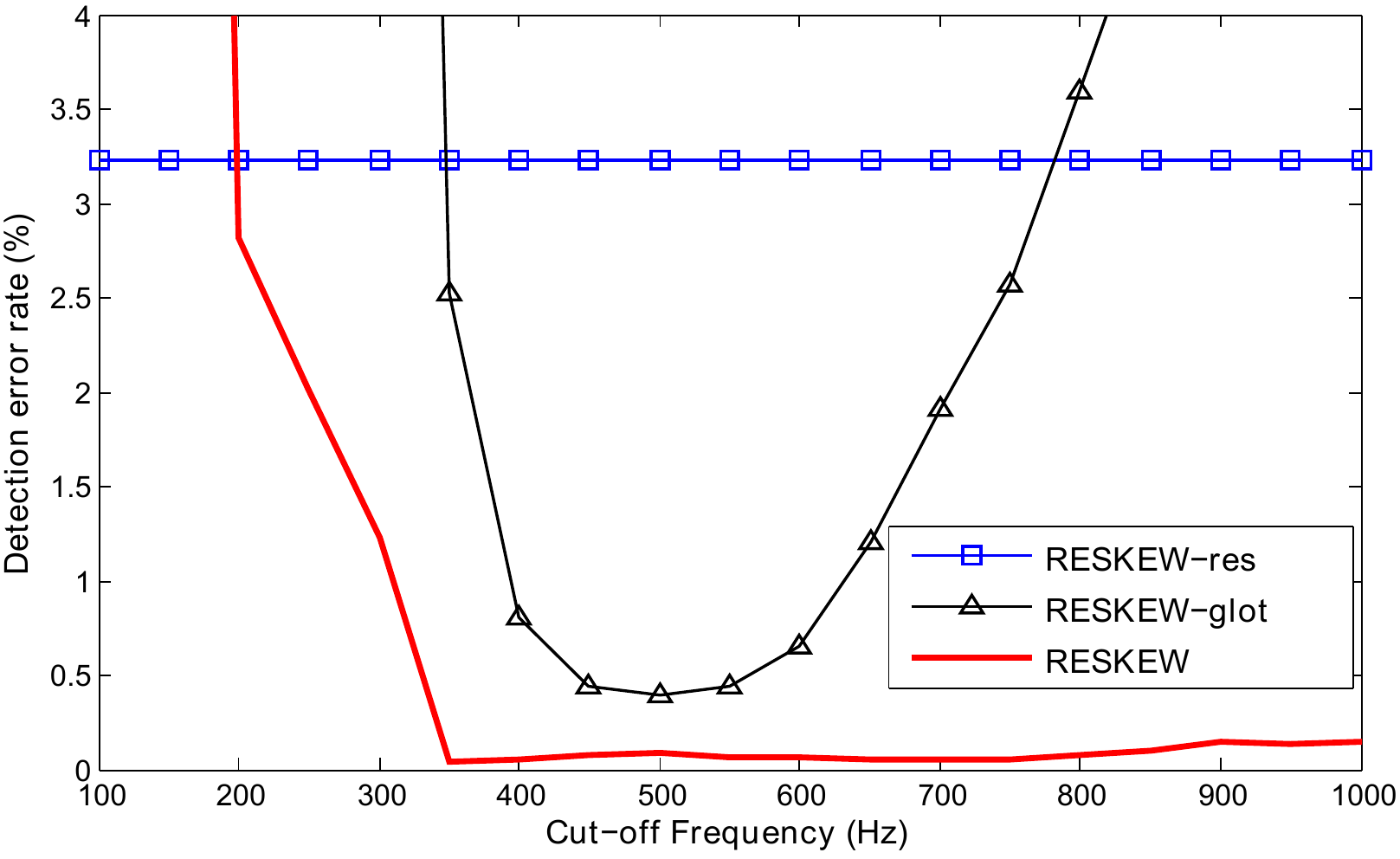}
  \vspace{-8pt}
  \caption{Impact of $F_c$ on the performance of the proposed methods.}
  \label{fig:CutOff}
  \vspace{-12pt}
\end{figure}

The influence of $F_c$ on the detection error rate is displayed in Figure \ref{fig:CutOff}. The performance of RESKEW-res is obviously independent of $F_c$, and reaches an error rate of 3.23\%. As for RESKEW-glot, it clearly goes by a minimum of about 0.4\% for $F_c$ around 500 Hz. Considering standard values for the first and glottal formants, our intuitive explanation for the choice of $F_c$ seems to be corroborated. Finally, RESKEW which combines both excitation signals turns out to have an efficient and stable performance in the range [350-800] Hz where it achieves error rates between 0.04 and 0.08\%. Nonetheless, in order to limit the impact of the degradation in noisy and reverberant environments, a low value of $F_c$ is preferred. We therefore fixed $F_c$ to 400 Hz in the rest of our experiments.

\subsection{Results in Clean Conditions}

\begin{figure}[!ht]
  \centering
  \includegraphics[width=0.4\textwidth]{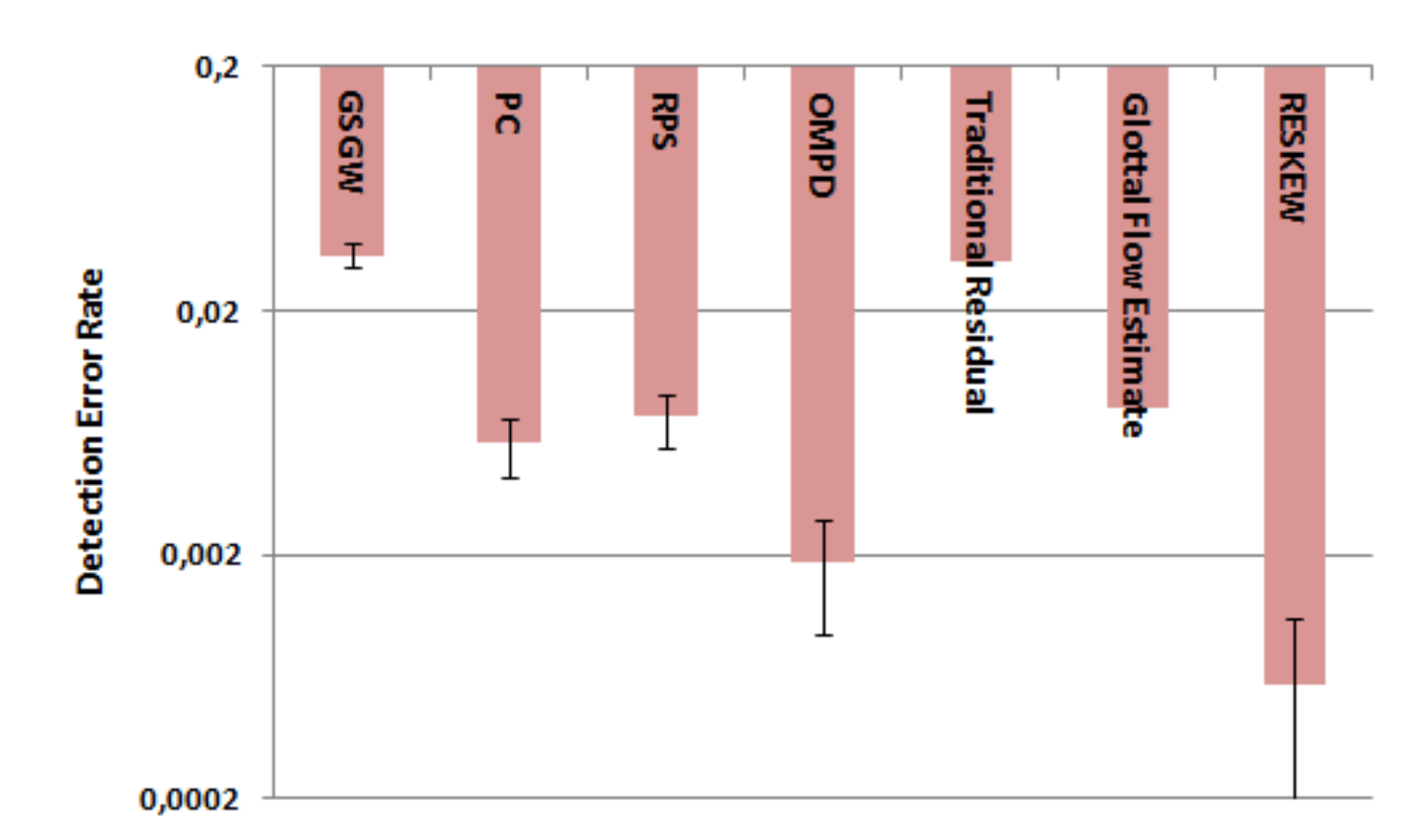}
  \vspace{-8pt}
  \caption{Averaged detection error rates obtained in clean conditions by the compared techniques, together with their 95\% confidence intervals.}
  \label{fig:CleanResults}
  \vspace{-12pt}
\end{figure}

Results we obtained on the original clean recordings are shown on a logarithmic scale in Figure \ref{fig:CleanResults}. With error rates of respectively 3.39 and 3.23\%, GSGW and RESKEW-res are clearly the worst techniques. RESKEW-glot, RPS and PC then give errors with the same order of magnitude: 0.81, 0.74 and 0.57\% respectively.  The second best method is OMPD (0.187\%) confirming its efficiency demonstrated in \cite{OMPD}. Finally, with an averaged detection error rate of only 0.0585\% (5 erroneous files out of the 8545), the proposed RESKEW algorithm clearly outperforms all other techniques. Note that these 5 errors are spread across four datasets. For an analysis of the speaker dependency of existing methods, the reader is referred to \cite{OMPD}. 

Regarding the computational load they impose, it is worth reminding that the four existing methods (see Section \ref{sec:Existing}) require the estimation of F0 and voicing decisions. This is achieved by the SRH approach with a RCT of 19.3\%. The resulting polarity detection techniques (GSGW, PC, RPS and OMPD) have total RCTs ranging between 30.5 and 34.1\%. In contrast, the proposed RESKEW method which is self-sufficient, reaches an averaged RCT of 12.3\% which makes it suitable for small devices or embedded systems. Note that since GSGW is the worst approach and led in \cite{OMPD} to prohibitive error rates in degraded conditions, its use will be discarded in the following.

\subsection{Results in Degraded Environments}

\begin{figure}[!ht]
  \centering
  \includegraphics[width=0.4\textwidth]{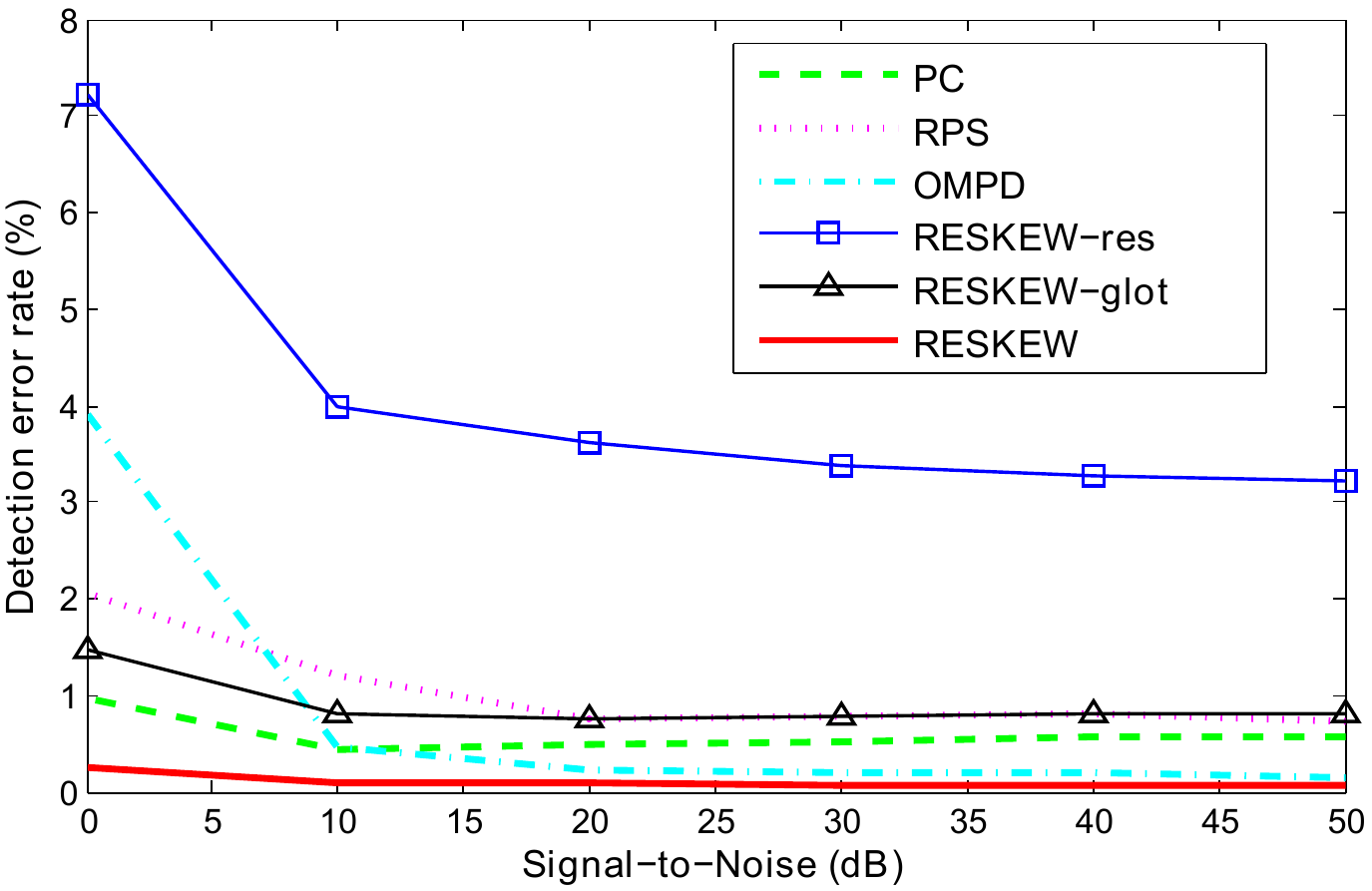}
  \vspace{-8pt}
  \caption{Robustness of the compared methods to an additive babble noise.}
  \label{fig:Babble}
  \vspace{-8pt}
\end{figure}

The performance with an additive babble noise is displayed in Figure \ref{fig:Babble}. It is observed that the most affected techniques at 0dB SNR are RESKEW-res and OMPD. Interestingly, the most robust method is RESKEW which maintains (mostly thanks to its glottal flow component) an almost constant efficieny independent of the SNR. This can be explained by the fact that the discontinuities in the glottal excitation at GCI locations are rather strong and spread and are preserved even in a noisy environment. As a consequence, their prevalent effect in the skewness calculation is maintained. Although results in a white noise are not presented here, conclusions that can be drawn are the same as with the babble noise.

\begin{figure}[!ht]
  \centering
  \includegraphics[width=0.4\textwidth]{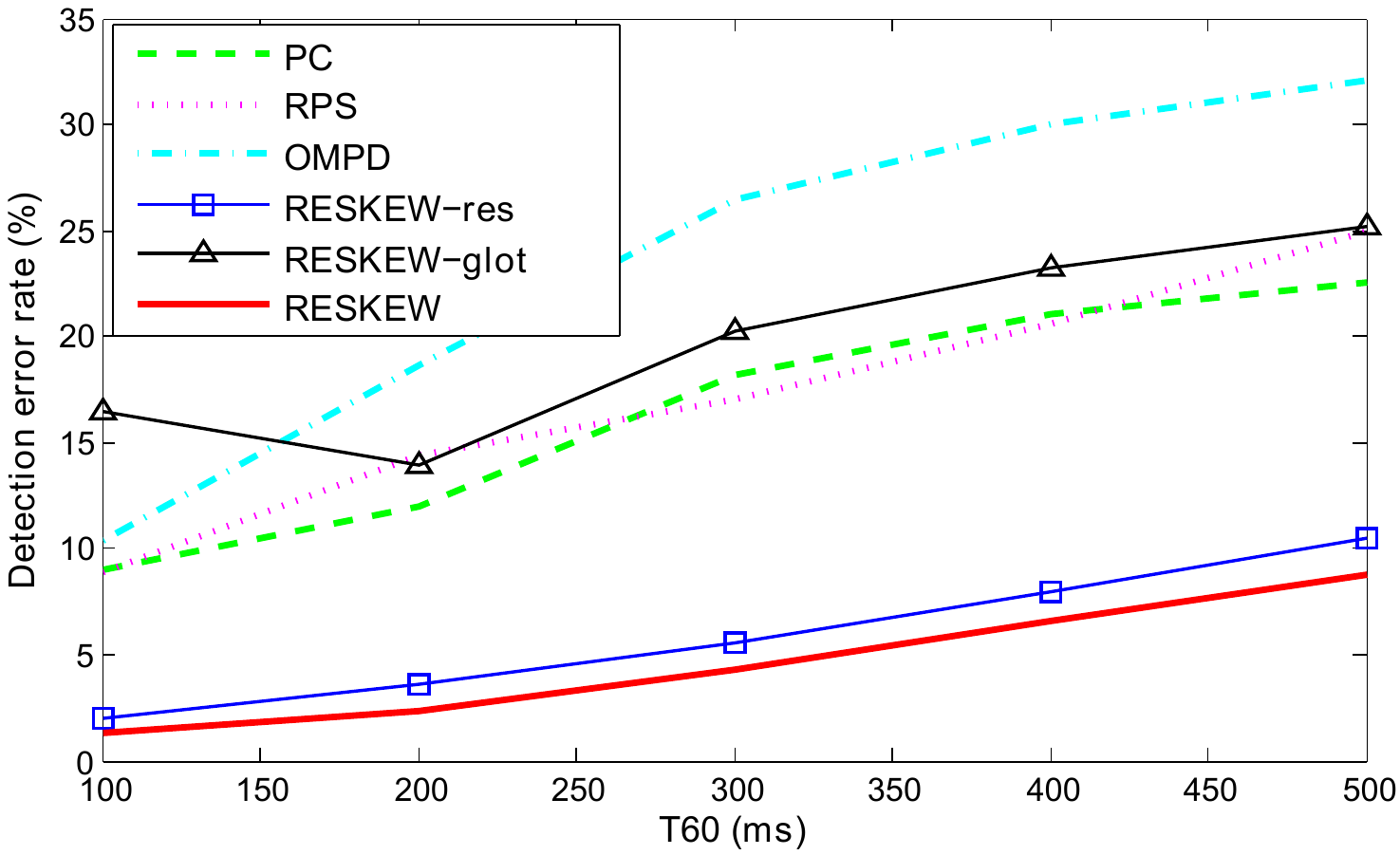}
  \vspace{-8pt}
  \caption{Robustness of the compared methods in a reverberant context.}
  \label{fig:Reverb}
  \vspace{-12pt}
\end{figure}

The impact of a reverberant environment is illustrated in Figure \ref{fig:Reverb}. It can be noticed that the degradation on PC, RPS, OMPD and RESKEW-glot is comparable. On the opposite, RESKEW and RESKEW-res can be distinguished by their remarkable robustness. Indeed, peaks in $g'(n)$ are rather spread out, while they are relatively sharp in $r(n)$ (see Figure \ref{fig:Example}). After multiple delayed reflections, the skewness of $g'(n)$ will be therefore much more affected (as large peaks are expected to overlap). An opposite effect occurs with an additive noise, where larger peaks will be more robust. Finally, it is worth emphasizing that RESKEW benefits from the robustness advantages of both its glottal source and residual signal components respectively in a noisy environment and in a reverberant context. It was shown to outperform all other approaches across all our experiments.

\section{Conclusion}
\label{sec:Conclu}
The goal of this paper was to propose a new technique of automatic polarity detection, called RESKEW. This method relied on two excitation signals: the LP residual, and a rough approximation of the glottal source. It was extensively compared to four state-of-the-art approaches on 10 large speech corpora. Thanks to its simplicity and the fact that it does not require further information such as voicing decision or F0, the proposed technique was shown to significantly reduce the computational load. In clean conditions, it reached an averaged error rate of 0.058\% against 0.187\% for the best existing method. Furthermore it was shown to be clearly the most robust in noisy conditions as well as in reverberant environments.


\ifCLASSOPTIONcaptionsoff
  \newpage
\fi

\end{document}